\documentclass[aps, singlecolumn, showpacs]{revtex4-2}
\usepackage{amsmath}
\begin{document}
\title{Topological photon, spin, and Euclidean group E(2)}
\author{S C Tiwari}
\affiliation{Department of Physics, Institute of Science, Banaras Hindu University, Varanasi 221005, and \\ Institute of Natural Philosophy \\
Varanasi India\\ Email address: $vns\_sctiwari@yahoo.com$ \\}
\begin{abstract}

The proposition that photon is a topological object (J. Math. Phys. 49, 032303, 2008) is given rigorous foundation based on pure vector field theory independent of the electromagnetic fields. Holomorphy of 4-dimensional space-time and the nature of topological obstructions are investigated to articulate singular vortex model for photon. The utility of the Euclidean symmetry group is discussed in detail for the proposed photon model. 
\end{abstract}

\maketitle

\section{\bf Introduction}

Planck and Einstein, the originators of the quantum oscillator and light quantum hypothesis respectively, in spite of their life-long efforts remained unsatisfied with the physical concept of photon \cite{1}. The name photon was given by Lewis to something different than Einstein's light quantum; see a review in \cite{2}. Lamb argues that the concept of photon is not only bad it is unnecessary or useless in quantum theory of radiation and quantum electrodynamics (QED) \cite{2}. It is true that the advances in quantum optics and precision tests of QED do not depend on the physical reality of photon, and the ubiquitous presence of the word "single photon" in the literature signifies only the operational and formal description of a calculational tool: a discrete excitation in a Fock state space of the radiation field, and/or the spatio-temporal mode amplitude in the modal space \cite{3}.

Even if there is apparently no compelling reason to seek physical reality of photon, it is also not satisfactory to accord absolute supremacy to empiricism and operationalism in physics. There are two important reasons to resolve the enigma of photon: foundations of quantum theory were laid by a new concept on photon in 1900, therefore, a physical model of photon has a potential to give new direction to the foundations of quantum theory, and secondly, it will add to the current efforts on quantum information science that utilize photon. We believe there is a need to depart from the historical and conventional path based on the electromagnetic fields and Maxwell equations for the description of photon to break the stagnancy on this issue \cite{4,5,6,7,8}. The new pathway may also lead to new physics. In this paper we explore the idea of topological photon \cite{8} in the perspective of group theoretic approach based on the Euclidean group E(2) in 2D. Though 2D plane plays fundamental role in the topological model of photon, the significance of the group E(2) and various facets of it, for example, the central extension $\bar{E}(2)$ and the covering group $\tilde{E}(2)$, were not realized at that time \cite{8}. It is also necessary to free the photon model from any imprint of the electromagnetic fields, i. e. a pure vector field description. In this connection Oppenheimer's "extreme light quantum theory" \cite{9} and the holomorphy of vector fields \cite{10} are re-visited. Once we postulate vector field as a fundamental photon field the question of gauge symmetry becomes redundant and a new approach is needed to study the E(2) like little group of Wigner in connection with photon spin  \cite{11}.

In the next section a thorough study is presented on the holomorphy of 4D space-time, pure vector field and field equations, and the question of the over-determination in Maxwell field theory and the establishment of the pure vector field equations for photon. Topological defects like dislocations, disclinations and singular vortex lines are explored in Section III for representing photon. Section IV is devoted to study the use of the Euclidean group in 2D in Wigner's little group for massless particles and the Aharonov-Bohm effect to gain new insights on the utility of the group E(2) for the vortex model of photon. The last section summarizes the new results and future outlook.

\section{\bf Vector field and photon model}

A brief review on the historical and conceptual aspects of photon physics is given in \cite{4} and in Chapter 2 of \cite{12}. In the proposed model of photon two important physical attributes are assumed: masslessness and spin angular momentum equal to $\hbar$. For a massless particle of any spin (neutrino, photon, graviton) relativistic space-time symmetries lead to the existence of only two helicity states. It has to be pointed out that experiments show that neutrino is not massless, therefore, massless neutrino may be treated as a mathematical object. In contrast, upper limit on photon mass from various experiments is consistent with massless photon.  Spin one of photon indicates that it is a vector field, spin half neutrino is a spinor field, and spin two graviton is a second rank tensor field. In \cite{8} photon is described by ${\bf E} =0$ and ${\bf B}=0$, and the electric and magnetic fields are defined in terms of the electromagnetic potential $({\bf A}, \Phi)$. In addition, the Lorentz gauge condition is assumed to describe the photon. In spite of a radically different hypothesis of assuming zero electromagnetic field for photon this assumption implies that the electromagnetic field is fundamental. Here we propose a pure vector field model independent of any recourse to the electromagnetic field. To distinguish this vector field from the electromagnetic potentials we denote it by $\mathcal{A}^\mu$. 

The importance of angular momentum of light for photon recognized by Oppenheimer \cite{9} in the vector field description unconnected with the Maxwell equations led to a self-dual six-vector field equation. This formulation could not be extended to incorporate the interaction with charges. However, it would not be a serious drawback if only a single isolated photon is considered. Unfortunately, Oppenhiemer's extreme light quantum theory does not resolve the fundamental questions related with the spin, wave-particle duality and physical reality of photon, and introduces more field variables as compared to a 4-vector field required for spin one photon. The notable aspect of Oppenheimer's approach is that photon divorced from Maxwell equations could be conceived by him and the role of photon spin was given importance when most of the physicists including Einstein did not attach much significance to photon spin \cite{1,4,5}.

In an elegant mathematical approach Salingaros \cite{10} studies  the holomorphic properties of 4D space-time using Clifford algebras. Note that in 2D the condition of holomorphy gives the Cauchy-Riemann equations. The definition and essential technicalities related with the "vee" $(\vee)$ product and the Cartan exterior product are given in the Appendix. In 4D space-time, holomorphic condition for a vector field $a$ 
\begin{equation}
D~\vee ~a=0
\end{equation}
implies a set of seven equations using the expression
\begin{equation}
D~\vee~a ={\bf \nabla}.{\bf a} +\partial^4 a^4 +(-\partial^4 {\bf a} - {\bf \nabla} a^4) \wedge \sigma^4 + \omega^3~ \vee ~[{\bf \nabla} \times {\bf a}]
\end{equation}
Equating the coefficients of each basis form in (2) equal to zero for the condition (1) to be satisfied we get the set of seven equations
\begin{equation}
{\bf \nabla}.{\bf a} +\partial^4 a^4=0
\end{equation}
\begin{equation}
\partial^4 {\bf a} + {\bf \nabla} a^4=0
\end{equation}
\begin{equation}
{\bf \nabla} \times {\bf a}=0
\end{equation}
If the vector field $a$ is identified with the 4-vector electromagnetic potential then Eq.(3) corresponds to the Lorentz gauge condition and Eq.(4) and Eq.(5) represent ${\bf E} =0$ and ${\bf B}=0$ respectively. 

For a tensor field of type 2 the holomorphic condition is equivalent to Maxwell field equations without sources. Salingaros considers this result to be of great physical import: this derivation of Maxwell equations is distinct from just the re-writing of the Maxwell equations in the form of Clifford algebra and implies the holomorphic structure of 4D space-time. 

The proposition that ${\bf E} ={\bf B}=0$ for photon and the use of the Lorentz gauge condition for photon field description \cite{8} would be compatible with the Eqns.(3), (4), and (5) representing the holomorphic structure of 4D space-time via the vector field.  We may use this analogy to justify our proposition in \cite{8}. However, we look afresh on these equations (3), (4), and (5) without invoking the electromagnetic field analogy. The set of equations (3)-(4)-(5) overdetermine the vector field that has only four components. If there are only four equations for the vector field the question arises whether we choose the pair Eq.(3) and Eq.(4) or the pair Eq.(3) and Eq.(5). This is a nontrivial question; in fact, we recall that eight Maxwell field equations also overdetermine the six field variables $({\bf E},~{\bf B})$ and there exist differing prescriptions for addressing this problem. To gain insights on this issue we make a digression and discuss Maxwell field equations in this context.

Source-free vacuum Maxwell field equations are first-order coupled partial differentail equations for $({\bf E},~{\bf B})$ in the 4D space-time variables 
\begin{equation}
{\bf \nabla}.{\bf E}=0
\end{equation}
\begin{equation}
{\bf \nabla}.{\bf B}=0
\end{equation}
\begin{equation}
{\bf \nabla} \times {\bf E} =-\frac{1}{c}\frac{\partial \bf B}{\partial t}
\end{equation}
\begin{equation}
{\bf \nabla} \times {\bf B} =\frac{1}{c} \frac{\partial \bf E}{\partial t}
\end{equation}
If we take the divergence of Eq.(9) and use the relation of vector calculus that the divergence of a curl vanishes then for time-dependent fields Eq.(6) follows immediately; similarly Eq.(7) follows from Eq.(8). Thus, the divergence equations (6) and (7) are not independent field equations, and we have six equations (8) and (9) for six field components $(E^i,~B^i); i=1,2,3$. For the description of the electromagnetic waves, the vector wave equations for electric and magnetic fields are obtained taking the curl of Eqns (8) and (9); the divergence equations (6) and (7) as subsidiary conditions define the transversality of the waves. 

In the literature, Schroedinger look-a-like form of Maxwell equations is also discussed introducing a complex vector field
\begin{equation}
{\bf \Psi} = {\bf E} + i {\bf B} 
\end{equation}
Multiplying Eq.(9) by $i$ and adding Eq.(8) and (9) one can re-write the Maxwell field equations as
\begin{equation}
-c {\bf S}.{\bf p} {\bf \Psi} =i \hbar \frac{\partial \bf \Psi}{\partial t}
\end{equation}
Here, $(S_j)_{kl} = i \epsilon_{jkl}$ satisfies the angular momentum Lie algebra of the group SO(3) $[S_j, ~S_k] = -i \epsilon_{jkl} S_l$. Good has discussed this formalism in analogy to Dirac relativistic wave equation for electron to interpret photon \cite{13} giving references to earlier works; however, his remark that this equation was discussed by Oppenheimer \cite{9} without identifying the real and imaginary parts of ${\bf \Psi}$ with electric and magnetic fields must be understood with caution because Oppenheimer's six-vector does not depend on the connection with the electromagnetic fields. The divergence equations could be re-written as 
\begin{equation}
{\bf \nabla}.{\bf \Psi} =0
\end{equation}
and it is just a subsidiary constraint equation not an independent field equation \cite{13}.

In the covariant Lagrangian formalism using variational principle an intriguing and interesting consequence is that the Maxwell field equations 
(6) and (9) arise as equations of motion, and equations (7) and (8) follow from the definition of the electromagnetic field tensor $F_{\mu\nu} =\partial_\mu A_\nu - \partial_\nu A_\mu$. The Lagrangian density for Maxwell field $\mathcal{L} \propto F^{\mu\nu} F_{\mu\nu}$ taking the variational field variable $A_\mu$ leads to the equation of motion
\begin{equation}
\partial^\mu F_{\mu\nu} =0
\end{equation}
The other equation
\begin{equation}
\partial_\mu F_{\nu \lambda} +\partial_\nu F_{\lambda\mu} +\partial_\lambda F_{\mu \nu}=0
\end{equation}
is called the Bianchi identity in the geometric formalism of the gauge field theory. The important point to note is that for one field $F_{\mu\nu}$ there is just one field equation (13) and there is no problem of overdetermination. It would appear that for electric and magnetic fields four equations (6) and (9) corresponding to Eq.(13) are fundamental; note that these two equations are necessary in the presence of sources, i. e. charge density and current density. Experimental Faraday's law (8) is contained in the Bianchi identity (14).

Let us re-consider Salingaros approach for the derivation of Maxwell equations. Duality theorem is used to get a canonical decompostion of second rank tensor in 4D space-time into 3D vector fields, and the D-derivative in Eq.(1) is
\begin{equation}
D= -\partial^i \sigma^i +\partial^4 \sigma^4
\end{equation}
Similar to Eq.(2) for vector field we get the D-derivative of $F$
\begin{equation}
D~\vee ~F ={\bf \nabla} \times {\bf B} -\partial^4 {\bf E} + {\bf \nabla} .{\bf E} ~\sigma^4 +\omega^4 \vee [-\partial^4 {\bf B} -{\bf \nabla} \times {\bf E} +{\bf \nabla}.{\bf B} ~\sigma^4]
\end{equation}
The holomorphic condition implies setting the coefficients of the basis forms in Eq.(16) equal to zero. The last square-bracketed term in Eq.(16) appears with the 4-volume form $\omega^4 = \sigma^1 \wedge  \sigma^2 \wedge  \sigma^3 \wedge  \sigma^4 $, and it is from this that the  Bianchi identity (14) or the pair (7)-(8) of Maxwell equations is obtained.

If we extend this logic to Eq.(2) then Eq.(5) obtained from the last term with the volume form could be viewed as an analogue of Bianchi identity, and the pair of Eq.(3) and Eq.(4) would be the fundamental field equations for the 4-vector field $a^\mu$. Remarkably the role of Eq.(5) as a subsidiary condition is important for relativistic covariance. Assumed 4-vector nature of $a_\mu$ in 4D space-time implies that Eq.(3) is Lorentz covariant transforming as Lorentz scalar. However, Eq.(4) does not satisfy relativistic invariance under the Lorentz transformation. For manifest Lorentz covariance both Eq.(4) and Eq.(5) are necessary: though Eq.(5) can be derived from Eq.(4) taking its curl and assuming time-dependent $a_\mu$, the identity (5) is necessary for Lorentz invariance.

In the light of above insights the vector field description for photon is proposed to be based on the fundamental field equations analogous to (3) and (4) satisfied by $\mathcal{A}_\mu$, and that to Eq.(5) as a constraint condition necessary for Lorentz covariance. In the formalism based on the Clifford algebras as well as in Weyl geometry the vector field has natural dimension of $(length)^{-1}$. 

\section{\bf Vector waves carrying topological obstruction }

The idea of a topological photon \cite{8} seems natural from mathematical point of view for the following reasons. Broadly, geometry and topology refer to local and continuous, and global and discrete/discontinuous aspects respectively. A simple example of interesting geometric property is that of a 2D surface of a 2-sphere, $S^2$. A vector parallel transported on a closed curve on the surface of $S^2$ acquires a change in its direction preserving the magnitude; it is termed holonomy in mathematics literature.  Geometric phases in optics based on the polarization state-space of the Poincare sphere or the spin-redirection phase on the momentum or wave vector space are the manifestations of the direction holonomy on the 2-sphere. On the other hand, a Euclidean 2D plane with a single point removed, i. e. a punctured plane $R^2 -\{ 0\}$, has nontrivial topology. A point singular vortex in fluid dynamics is one of the well-known examples. If a line, say along z-axis, is removed in 3D Euclidean space then nontrivial topology is that of the product of a circle and a 2D plane. Vortex-line in fluid flow, for example, in a superfluid and the Aharonov-Bohm effect represent important examples of this topological property. A point charge of electron may be viewed as a topological obstruction in 3D space: total electric flux through any closed surface irrespective of its shape and size surrounding the electron is constant. In semiclassical Bohr-Sommerfeld quantization, orbital manifold (orbifold) is a topological obstruction. Mathematical theory of differential forms and topology in terms of de Rham periods is very useful in understanding topological obstructions. A brief account of de Rham periods is given in \cite{8}. For a nice physics-oriented approach and references we refer to Post's monograph \cite{14}. 

Abstract de Rham period rendition of Bohr-Sommerfeld quantization has been utilized to interpret photon spin as a topological defect \cite{8}. Elementary arguments to explain spin angular momentum in terms of dynamical fields require the transverse fields: assuming the direction of propagation to be z-axis there must exist nonvanishing fields in the xy-plane. Phase singularities, dislocations, and disclinations in fields are examples of widely discussed topological objects in waves. While phase singularities and dislocations could arise in scalar waves, disclinations and polarization singularities exist in vector waves \cite{15}. The question is what kind of fields represent photon.

Experiments show particulate discrete nature as well as continuous wave property of photon. In the standard Copenhagen interpretation of Bohr one associates wave-particle duality and the complementarity principle with this phenomena \cite{16}. However, de Broglie believed in wave and particle interpretation \cite{16,17}. Topological model of photon naturally accounts for particle aspect in terms of spin as a topological defect, and supports de Broglie's wave and particle interpretation. Spin one of photon demands vector wave bearing the topological obstruction for its description. Since we propose vector waves independent of the electromagnetic waves for photon we term them as $\mathcal{A}$-vector waves and use ${\bf E}$-vector waves for the electromagnetic waves. The proposed photon field equations are of the form (3), (4), and (5) written explicitly below for $\mathcal{A}_\mu$
\begin{equation}
{\bf \nabla}.{\bf \mathcal{A}} +\frac{1}{c} \frac{\partial \mathcal{\phi}}{\partial t} =0
\end{equation}
\begin{equation}
{\bf \nabla} \mathcal{\phi} +\frac{1}{c} \frac{\partial {\bf \mathcal{A}}}{\partial t} =0
\end{equation}
\begin{equation}
{\bf \nabla} \times {\bf \mathcal{A}} =0
\end{equation}

Vector waves for $\mathcal{A}_\mu$ can be introduced in two different ways. Taking the curl of (19) and using Eq.(17) and Eq.(18) we obtain the vector wave equation
\begin{equation}
\nabla^2 {\bf \mathcal{A}} = \frac{1}{c^2} \frac{\partial^2 {\bf \mathcal{ A}}}{\partial t^2}
\end{equation}
and taking the divergence of (18) and using Eq.(17) we get
\begin{equation}
\nabla^2 \mathcal{\phi} = \frac{1}{c^2} \frac{\partial^2 \mathcal{\phi}}{\partial t^2}
\end{equation}
In this way we get vector wave equations (20) and (21) for the 4-vector field $\mathcal{A}_\mu$.

Alternative approach could be to introduce a scalar field $\eta$ assuming
\begin{equation}
{\bf \mathcal{A}} ={\bf \nabla} \eta ~ ;~~ \mathcal{\phi} =-\frac{1}{c} \frac{\partial \eta}{\partial t}
\end{equation}
It is easy to show that Eq.(18) and Eq.(19) are identically satisfied, and Eq.(17) leads to the scalar wave equation
\begin{equation}
\nabla^2 \eta = \frac{1}{c^2} \frac{\partial^2 \eta}{\partial t^2}
\end{equation}
In this case the vector wave properties are derived using the solution of scalar wave equation (23) and substituting that in Eq.(22).

A large class of solutions of scalar and vector wave equations is known that is discussed in the textbooks and specialized vast literature. Here our aim is to focus on the nontrivial topological objects-bearing waves. Three types of solutions are examined below.

{\bf S1: Pure screw disclination}

A concise and nice exposition on the terminology and the nature of the topological defects in the wave phenomena is given in \cite{15}. In scalar optics the 3D diffraction pattern shows dark lines where the amplitude of the scalar field vanishes. In analogy to crystal dislocations the lines are termed wave dislocations. In the complex representation of a scalar field
\begin{equation}
\Psi ({\bf r}, t) = \rho  ({\bf r}, t) e^{i \chi  ({\bf r}, t)}
\end{equation}
wave dislocation is defined as a singularity where
\begin{equation}
 \rho  ({\bf r}, t) =0
\end{equation}
and the phase $ \chi  ({\bf r}, t) $ is indeterminate. Assuming z-axis as the propagation direction a typical screw dislocation is obtained for
\begin{equation}
\Psi = k (ax +i y) e^{i \chi(z,t)}
\end{equation}
\begin{equation}
\chi(z,t)= k z - \omega t
\end{equation}
where $a >0; ~ \omega = k c$. Exact solution for edge dislocation is given by
\begin{equation}
\Psi = k (a x + k x^2 +i z) e^{i(kz-\omega t)}
\end{equation}
Eq.(28) represents two edge dislocations in the xy-plane at $x=0$ and $x=-a/k$ parallel to the y-axis. The dislocation at $x=0$ is intrinsic while making the real number 'a' to be very large the second dislocation can be moved far away. The condition (24) implies that both real and imaginary parts of $\Psi$ have to vanish resulting into a line of the intersection of two surfaces that defines the dislocation.

Polarization singularities in ${\bf E}$-vector waves are termed disclinations in analogy to the defects in liquid crystals. A line singularity in the direction of the transverse field represents a typical disclination. For ${\bf E}$-vector waves ${\bf E}$ and ${\bf B}$ satisfy the vector wave equations and the divergence equations (6) and (7) impose transversality condition. Each component of electric field vectors satisfies a scalar wave equation, and for the assumed monochromatic waves, Eq.(8) is used to determine the magnetic field \cite{15}. Assuming z-axis as the direction of propagation the disclination for ${\bf E}$- waves is defined by 
\begin{equation}
E_x({\bf r},t) =B_x({\bf r},t)  =0
\end{equation}
To obtain disclination solutions Nye adopts scalar dislocation solutions and constructs ${\bf E}$ and ${\bf B}$ fields from them. One of the general solutions discussed by him is given by 
\begin{equation}
E_x = k r^\prime e^{i (\theta^\prime +\chi)}
\end{equation}
\begin{equation}
E_y =i  k r^\prime e^{i (\theta^\prime +\chi)}
\end{equation}
\begin{equation}
E_z =i (a- \cos \delta) e^{i \chi}
\end{equation}
Here the rotated coordinates in the stretched space $x^\prime =a x$ are introduced
\begin{equation}
r^\prime = {({x^\prime}^ 2 +{y^\prime}^ 2)}^{\frac{1}{2}}; ~~ x^\prime =a x = r^\prime \cos \theta^\prime ~~ y^\prime =r^\prime \sin \theta^\prime
\end{equation}
and $\delta$ denotes the angle between $z^\prime$-axis and the wave vector of the perturbing plane wave. Eq.(31) for $E_y$ is obtained from $E_x$ introducing a phase difference of $\pi /2$, and Eq.(32) is derived using the divergence equation (6). A number of interesting consequences for the monochromatic electromagnetic waves carrying disclinations are discussed in \cite{15}. One of the important roles of topological defects is the emergence of a double helix structure in which disclination in ${\bf E}$ and disclination in ${\bf B}$ wind each other with a spacing of the order of the wavelength of the wave divided by $2 \pi$.

In the present paper we consider one of the special cases that of pure disclination for $\mathcal{A}$-vector waves. In this case $\delta =0,~a=1$, and $\theta$ is just the polar angle in the xy-plane. For ${\bf E}$-vector waves Eq.(32) shows that $E_z$ vanishes. In contrast to this, for $\mathcal{A}$-vector wave two nonvanishing fields $\mathcal{A}_z$ and $\mathcal{\phi}$ exist besides the transverse fields. In \cite{8} we assumed the transverse fields $\mathcal{A}_x ,\mathcal{A}_y$ of exactly the same form as that of (30) and (31) for $E_x, E_y$ respectively
\begin{equation}
\mathcal{A}_x = k r^\prime e^{i (\theta^\prime +\chi)}
\end{equation}
\begin{equation}
\mathcal{A}_y =i  k r^\prime e^{i (\theta^\prime +\chi)}
\end{equation}
The azimuth of the transverse field in the xy-plane 
\begin{equation}
\beta = -(\chi +\theta)+ 2 n \pi
\end{equation}
was re-interpreted to introduce topological defect in time in \cite{8}. Nye \cite{15} interprets the case at z=0 as a rigid rotation of the whole pattern with angular velocity $\omega/2$, and for fixed t the pattern twists about the z-axis by $\pi$ in one wavelength. We re-examined the case z=0
\begin{equation}
\beta - \theta  = - 2(\theta -\frac{\omega t}{2}) +2 n \pi
\end{equation}
and interpreted screw disclination of phase change $\pi$ as a tifold \cite{8}.

Consistency of the solution with the set of the field equations (17), (18), and (19) offers a new insight into the topological defect in time. Eq.(17) for $\mathcal{A}_z, ~~\mathcal{\phi}$ reads
\begin{equation}
\frac{\partial  \mathcal{A}_z}{\partial z} + \frac{1}{c} \frac{\partial \mathcal{\phi}}{\partial t} =0
\end{equation}
A longitudinal propagating plane wave solution of Eq.(38) is assumed
\begin{equation}
\mathcal{A}_z = e^{i \chi} , ~~\mathcal{\phi} = e^{i\chi}
\end{equation}
Eq.(18) in this case shows that 
\begin{equation}
\frac{\partial  \mathcal{A}_x}{\partial t} =\frac{\partial  \mathcal{A}_y}{\partial t} =0
\end{equation}
Therefore, the assumed solutions (34) and (35) for $\mathcal{A}_x$ and $\mathcal{A}_y$ respectively are inconsistent with (40). 

Mathematical consistency can be achieved in two ways. First, let us assume following time-independent solution for transverse fields
\begin{equation}
 \mathcal{A}_x = k (x+i y)
\end{equation}
\begin{equation}
 \mathcal{A}_y = i k (x+i y)
\end{equation}
This solution together with (39) is a consistent solution of the set of field equations (17), (18), and (19). Physical interpretation follows from the recognition that both $ \mathcal{A}_x , ~ \mathcal{A}_y$ possess intrinsic dislocations : amplitude $r$ vanishes and the phase $\arctan \frac{y}{x}$ is indeterminate. 

More interesting possibility arises if in analogy to orbifold we postulate nontrivial 1D directed punctured line $R^1 - \{0\}$ for time. Now, time-independence of $ \mathcal{A}_x , ~ \mathcal{A}_y$ holds in 2+1D $R^2 -\{0\} \oplus R^1 - \{0\}$. Screw disclination as a tifold based on Eq.(37) can be envisaged in this 2+1D space-time, and internal time in the xy-plane could be associated in harmony with the frequency of the propagating wave.

{\bf S2: A pair of dislocations}

The topology of $\mathcal{A}$-vector waves in a different picture arises beginning with the scalar field $\eta$. Assuming scalar wave dislocation defined by
\begin{equation} 
\eta =(x+i y) e^{i \chi}
\end{equation}
for the solution of Eq.(21) the vector field components obtained using Eq.(22) are given by
\begin{equation}
\mathcal{A}_x =e^{i \chi} ; ~~ \mathcal{A}_y =i e^{i \chi}
\end{equation}
\begin{equation}
\mathcal{A}_z= i k (x+i y)e^{i \chi} 
\end{equation}
\begin{equation}
\mathcal{\phi} = i k (x+i y)e^{i \chi}
\end{equation}
Expressions (44) show that the $\mathcal{A}$-vector wave is ciscularly polarized, and expressions (45) and (46) represent intrinsic screw dislocations in $\mathcal{A}_z$ and $\mathcal{\phi}$ respectively. 

{\bf S3: Vortices - New Perspective}

In order to appreciate the new perspective on the vortices put forward here, we make two observations. In the 19-th century, the dynamical theory of fluids was in a very advanced state, and historically the fluid dynamical analogy played crucial and constructive role in the development of the Maxwell's theory of the electromagnetic fields. Analogy to singular and continuous vortices in fluids also motivated vortex models for aether, electromagnetic fields and matter \cite{18}. Past few decades have witnessed re-newed interest and immense advances in the experimental and theoretical investigations in the field of optical vortices. Let us take a simple example. The scalar wave dislocations defined by Eq.(14) and Eq.(15) could be re-interpreted as optical vortices using the terminology of fluid flow. A monochromatic cylindrical light beam in the paraxial approximation can be represented by a complex scalar field of the following form
\begin{equation}
E= E_0(r,z)~e^{i(\Phi +k z +l \theta -\omega t)}
\end{equation}
At time $t=0$ the gradient of the phase in (47) gives
\begin{equation}
{\bf \nabla} (k z +l \theta +\Phi) = k \hat{z} +\frac{l}{r} \hat{\theta} +{\bf \nabla} \Phi
\end{equation}
Identifying the phase as the velocity potential for a fluid flow the phase singularity in (48) represents a vortex with topological charge $l$; for a nice review see \cite{19}, and a short discussion can also be found in \cite{4}.

The most important difference between the standard approach on vortices in optics and electromagnetic fields and the present one is that we consider vortices in pure vector field $\mathcal{A}$. Let us assume following solution for the scalar wave equation (23) 
\begin{equation}
\eta ({\bf r}, t) = \eta_0(x,y) ~e^{i(k z -\omega t)}
\end{equation}
Assuming $k^2 c^2 =\omega^2$, we have
\begin{equation}
\frac{\partial^2 \eta}{\partial x^2}+\frac{\partial^2 \eta}{\partial y^2}=0
\end{equation}
Postulating the space-time continuum as a kind of fluid (aether!) we define a stream function $S(x,y)$ that is a perfect differential
\begin{equation}
dS= \mathcal{A}_y dx - \mathcal{A}_x dy
\end{equation}
where $\mathcal{A}_x =-\frac{\partial S}{\partial y},~\mathcal{A}_y =\frac{\partial S}{\partial x}$. Eq.(17) implies that constant $\eta_0$ and constant $S$ intersect orthogonally. A vortex solution using polar coordinates is obtained for $\eta_0 = \theta , S =-\ln r$. We can calculate the vector field components to be
\begin{equation}
\mathcal{A}_r =0
\end{equation}
\begin{equation}
\mathcal{A}_\theta = \frac{1}{r}
\end{equation}
Treating (52) and (53) as momentum field components it follows that the flow is in the concentric circles around the origin as the radial momentum (52) is zero, and the singularity at $r=0$ in Eq.(53) defines a vortex filament.

To investigate the nature of the topological obstruction in more detail we adopt the abstract concept of de Rham periods. Let us first re-write (52)-(53) in the cartesian coordinates
\begin{equation}
\mathcal{A}_x =-\frac{y}{x^2+y^2}
\end{equation}
\begin{equation}
\mathcal{A}_y =\frac{x}{x^2+y^2}
\end{equation}
Recalling the language of differential forms \cite{8} the Hodge decomposition of a 1-form $A$ states
\begin{equation}
A = df +\partial g +\gamma
\end{equation}
where first term on the RHS of (56) is an exact form, the second term is the closed form, and the last one is the harmonic form. The integral of (56) around a closed loop gives the de Rham periods
\begin{equation}
\oint A =\oint df+\oint \partial g +\oint \gamma
\end{equation}

In a simpler language of vector calculus in the Euclidean geometry $R^3$, the gradient of a scalar with vanishing curl, and the curl of a vector with zero divergence correspond to exact and closed forms respectively. Harmonic form is an interesting object that corresponds to a vector field whose divergence and curl both vanish. In a nontrivial geometry, the harmonic form may contain information about topological obstructions. The important example is that of a vector field defined by (54)-(55) that has zero divergence and zero curl in the whole domain of $R^3 -\{0\}$. The de Rham period for this harmonic form is
\begin{equation}
\oint \gamma = \oint \frac{y ~dx-x~dy}{x^2+y^2}= 2\pi n
\end{equation}
Since the transverse vector field components $\mathcal{A}_x,~\mathcal{A}_y$ in the fluid interpretation represent momentum or velocity fields, it is natural to relate the de Rham period (58) with the vorticity. To complete the concrete physical model of photon, the longitudinal vector field $\mathcal{A}_z$ describes a propagating wave using Eq.(22). It may be noted that Kiehn also suggests a photon as a propagating discontinuity or defect \cite{20}. However, Kiehn's model is based on the electromagnetic fields, the defect is defined such that "${\bf E}$ and ${\bf B}$ are both zero, and at the same time, ${\bf E}$ and ${\bf B}$ are not zero".

\section{\bf Euclidean group E(2) and photon}

The standard representation of a photon in QED/quantum optics \cite{3} is in terms of a plane wave defined by
\begin{equation}
A^\mu(x,k) = \epsilon^\mu (\frac{\hbar}{2\omega V})^\frac{1}{2} (e^{-i k x} +e^{+i k x})
\end{equation}
Here $\hbar k^\mu$ is a momentum 4-vector, and $\epsilon^\mu$ is a unit polarization vector satisfying the massless condition
\begin{equation}
k_\mu k^\mu =0
\end{equation}
and the transversality condition
\begin{equation}
\epsilon_\mu k^\mu =0
\end{equation}
The physical quantities, for example, energy, momentum, and angular momentum are obtained from the electromagnetic fields $({\bf E},{\bf B})$. The normalization constant in (59) can be chosen using the energy expression $U=\frac{1}{2} \int ({\bf E}.{\bf E} + {\bf B}. {\bf B})~ dV$, and equating it with $\hbar \omega$. In a simple picture, a photon is a discrete excitation in the Fock space with wave vector ${\bf k}$ and polarization index, say, $s$. Energy $\hbar \omega$ is independent of the polarization and the direction of ${\bf k}$. Intrinsic spin angular momentum is $\pm \frac{\hbar {\bf k}}{|{\bf k}|}$, and it lies along the direction of propagation.

One of the important approaches to quantum field theories is to consider the space-time symmetries of the inhomogeneous Lorentz group (the Poincare group). Wigner in 1939 gave a reasonably rigorous  theory \cite{21} for invariant linear manifolds of the quantum states in a Hilbert space  building on the earlier works of Majorana and Dirac. Unitary and irreducible unitary representations for the Poincare group on such invariant manifolds were discussed by him. Since space-time translation operators commute with each other, one can choose a coordinate system in a Hilbert space such that the wavefunctions contain momentum variables $p_\mu$. Further, in a homogeneous Lorentz group one can choose a uniformly moving coordinate system $K^\prime$  to have the same origin as that of the rest frame $K$ at $t=0$, and transform the direction of the uniform motion along, say, z-axis. In this construction, the two frames can differ only by rotation. Group space of the Lorentz transformation is shown to be doubly connected. 

Relativistic particles of non-zero rest mass and arbitrary spin can be represented using the Poincare space-time symmetries and Wigner's little group. For a massless particle there arise subtle aspects as one cannot have a rest frame in this case. For massive and massless particles the little groups are shown to be isomorphic to 3D rotation group SO(3) and 2D Euclidean group E(2) respectively \cite{21}. Kim and Wigner \cite{22} re-visited the little group for photons to gain new insights on the physical import of translations in E(2).  A cylindrical group isomorphic to E(2) was introduced by the authors and the translations were related with the gauge symmetry. Here, two interesting papers deserve to be mentioned published prior to \cite{22}. First one concerns a pedagogical discussion on Wigner's little group and gauge transformations \cite{23} based on Weinberg's arguments \cite{24}. Does spin of photon originate from the rotation in gauge space? This question was addressed in \cite{11}

Now, little group arises from the group of Lorentz transformations $L^\mu_\nu$ applied to a light-like 4-vector $k^\mu$
\begin{equation} 
L^\mu_\nu k^\nu = k^\mu
\end{equation}
For a single photon propagating along z-axis we have $k^1 =k^2 =0 ; ~k^3 =k, k^0 =k c$. The Lorentz transformation matrix in this case can be factored into a roatation matrix and translations in E(2): $R(\alpha) D(u,v)$. Explicitly, for the rotation parameter $\alpha$ and translation parameters $(u,v)$ for E(2) we have
\begin{equation}
L^\mu_\nu = \begin{bmatrix} \cos \alpha & \sin \alpha & 0 & 0 \\ -\sin \alpha & \cos \alpha & 0 & 0 \\ 0 & 0 & 1 & 0 \\ 0 & 0 & 0 & 1 \end{bmatrix} \begin{bmatrix} 1 & 0 & -u & u \\ 0 & 1 & -v & v \\ u & v & 1 -(u^2 +v^2)/2 & (u^2 +v^2)/2 \\  u & v & -(u^2 +v^2)/2 & 1 +(u^2 +v^2)/2 \end{bmatrix}
\end{equation}
Kim and Wigner \cite{22} note that the quadratic terms in the variables $(u,v)$ make it difficult to reduce the $4 \times 4$ matrix (63) into the block diagonal form. Imposing the Lorentz gauge condition for the vector potential, the $4 \times 4$ representation of the little group is reduced to the block diagonal form in which one gets the generators of rotation and E(2). 

The importance of E(2) as a little group for photon \cite{21,22,23}, and the speculation relating spin of photon with the rotation in gauge space \cite{11} need radical revision in the pure vector field approach proposed for photon in the present paper. To delineate novel elements that we introduce, first a brief discussion on the relevant ideas in \cite{11,24} is presented. Point field theory and photon as a Maxwell field constitute the basic premise in the conventional photon theory.  Weinberg \cite{24} studies the invariance of S-matrix, $S_{\pm 1}({\bf q},Q)$, considering the emission of a photon with momentum ${\bf q}$, and helicity $\pm 1$. Here, $Q$ denotes the momenta and helicities of all other particles taking part in the scattering process. Three important points are made by Weinberg. (i) For the correct Lorentz transformation of the S-matrix, it is necessary and sufficient that $S_\pm$ vanish when one of the $\epsilon_\pm^\mu$ is replaced with $q^\mu$. Note that Weinberg assumes polarization vector to be
 \begin{equation}
\epsilon^\mu_\pm = \frac{1}{\sqrt{2}} \begin{bmatrix} 1 \\ \pm i \\0 \\0 \end{bmatrix}
\end{equation}
(ii) The envisaged condition in (i)  could be interpreted as mass-shell gauge invariance since S-matrix is invariant under the gauge transformation
\begin{equation}
\epsilon_\pm^\mu (\hat{q}) \rightarrow \epsilon_\pm^\mu (\hat{q})+\lambda_\pm ({\bf q}) q^\mu
\end{equation}
where $\lambda_\pm ({\bf q})$ is an arbitrary function.
(iii) In the Appendix A, an important relation (A19) is derived for the Lorentz transformation $\Lambda^\mu_\nu \epsilon^\nu_\pm (\Lambda {\bf q})$ in which appears the gradient term arising from the translations $(u,v)$ of the little group.

Han and Kim \cite{23} have explained Weinberg's points in a simple manner. In a later paper \cite{11} an attempt is made to give a tentative picture of photon spin. Let us consider Eq.(63) for $\alpha =0$ and use the notation $D(\alpha,u,v)$ for $L^\mu_\nu$, then
\begin{equation}
D(0,u,v) =\begin{bmatrix} 1 & 0 & -u & u \\ 0 & 1 & -v & v \\  u & v & 1-\frac{(u^2 +v^2)}{2} & \frac{(u^2 +v^2)}{2} \\ u & v & -\frac{(u^2 +v^2)}{2} & 1+\frac{(u^2 +v^2)}{2} \end{bmatrix}
\end{equation}
The matrix (66) can be re-written as $D(0,u,v) = D_1(u) D_2(v)$ where
\begin{equation}
D_1(u) = \begin{bmatrix} 1 & 0 & -u & u \\ 0 & 1 & 0 & 0 \\  u & 0 & 1-\frac{u^2 }{2} & \frac{u^2}{2} \\ u & 0 & -\frac{u^2}{2} & 1+\frac{u^2}{2} \end{bmatrix}
\end{equation}
\begin{equation}
D_2(v) =  \begin{bmatrix} 1 & 0 & 0 & 0 \\ 0 & 1 & -v & v \\  0 & v & 1-\frac{v^2 }{2} & \frac{v^2}{2} \\ 0 & v & -\frac{v^2}{2} & 1+\frac{v^2}{2} \end{bmatrix}
\end{equation}
The generators of the Lie algebra e(2) are the translation operators $N_1, N_2$ and the rotation operator $J_3$. The three generators commute with
\begin{equation}
N^2 = N_1^2+N_2^2
\end{equation}
If the transformation matrix (66) operates on the polarization state (64) we get
\begin{equation}
\epsilon^\mu_\pm~ \rightarrow ~\frac{1}{\sqrt{2}} \begin{bmatrix} 1 \\ \pm i \\u \pm i v \\ u \pm iv \end{bmatrix}
\end{equation}
Authors \cite{11} note that the unusual transformation (70) needs explanation; the eigenvalue equation for (69)
\begin{equation}
N^2 \psi(u,v) = -(\frac{\partial^2 \psi}{\partial u^2} +\frac{\partial^2 \psi}{\partial v^2}) = b^2 \psi
\end{equation}
has non-singular solution in polar coordinates for eigenvalue $b^2 =0$ zero
\begin{equation}
\psi = U e^{\pm i \alpha} ; ~ U^2= u^2+v^2, \alpha = \arctan \frac{v}{u}
\end{equation}
This solution corresponds to (70). Photon spin is interpreted in terms of a rotation in 2D gauge space.

The reason for the need of gauge space in \cite{11} is that the electromagnetic fields are assumed to be real physical fields and the 4-vector potential is arbitrary up to a gauge transformation. The postulate that 4-vector field $\mathcal{A}^\mu$ is a fundamental physical field for a single photon makes the gauge symmetry and gauge space redundant. Analogy to fluid dynamics leads to identify $\mathcal{A}^\mu$ as velocity/momentum field, and envisage rotation in 2D physical space for photon spin. Transverse 2D plane would logically imply Euclidean symmetry group, but it cannot be the usual group E(2). Space translation symmetry in the transverse plane gives the momentum fields $(\mathcal{A}_x, \mathcal{A}_y)$; multiplying 4-vector field $\mathcal{A}^\mu$ having the geometric dimension of $length^{-1}$ by $\hbar$ we get the energy- momentum vector field. 

The solution (72) is, in fact, a trivial simple solution that satisfies any second order partial differential equation of two variables. Novel aspects could be attributed to it from the topological perspective. The use of cylindrical coordinates has in-built indeterminacy in defining the origin $O$: in cartesian coordinate system the origin is defined as $x=0, y=0; O(0,0)$, however, in polar coordinates the polar angle $\arctan \frac{0}{0}$ is not defined at the origin. The complex function (72) would represent a phase vortex of the kind (24) since at zero amplitude the phase is indeterminate. From the holomorphic point of view \cite{10}, the function (72) satisfies the Cauchy-Riemann equations that follow from the holomorphic condition in 2D. 

A curious interesting result is obtained if we use (72) for the solution of the Schroedinger wave equation for a free particle
\begin{equation}
\Psi_S(x,y,z,t) = (x+i y) e^{i(k z -\omega t)}
\end{equation}
and calculate the current density ${\bf J}_S =\frac{\hbar}{2 i m}( \Psi_S^* {\bf \nabla} \Psi_s - ({\bf \nabla} \Psi_S^*) \Psi_S)$. In the Copenhagen interpretation this current density defines the probability current density. However, in the hydrodynamical interpretation of Madelung \cite{16} this represents the fluid flow current density, and one can define "the velocity field" ${\bf v}_S =\frac{ {\bf J}_S}{\Psi_S^* \Psi_S}$. Using (73) the velocity field is obtained to be
\begin{equation}
{\bf v}_S = \frac{\hbar}{m} (\frac{-y \hat{x} + x \hat{y}}{x^2+y^2} + k \hat{z})
\end{equation}
The velocity field (74) contains the vortex field (54)-(55). Obviously the photon vector field is not the Schroedinger wavefunction, however the point that we wish to underline is that the topological defect represented by vortex field (54)-(55) may be more general than so far realized in the literature.

The vector field (54)-(55) is discussed in the Appendix C of \cite{25} as an example of a nonexact gauge field. It is noted that the curl of the vector field is zero in $R^3 -\{0\}$, and the loop integral of the vector field can be calculated using the polar coordinates $\oint d\theta =2\pi$; the gauge vector field cannot be a gradient field globally. Two additional points need to be emphasized. The vector field has zero divergence in addition to the vanishing curl, and, therefore, it is a harmonic form. Transforming to polar coordinates, one has to restrict the polar angle to $[0, 2\pi)$, otherwise one would get the value of the integral to be $2 n \pi$ since the real line $R^1$ is a universal covering of the circle $S^1$. 

The vector fields $\mathcal{A}_x, \mathcal{A}_y$ given by Eqs.(54-55) or the velocity fields $v_x, v_y$ in Eq.(74) in the punctured transverse 2D plane need the symmetry group other than the usual Euclidean group E(2). Let us examine a mathematical model of the Aharonov-Bohm effect \cite{26}. The motion of an electron in the absence of a  magnetic flux line has the symmetry group E(2); however, non-zero flux along the z-axis requires a q-fold covering group $\bar{E}(2)$. Martin \cite{26} presents both state space and self-adjoint operators to discuss the q-fold covering group $\bar{E}(2)$. The infinitesimal generator of rotation is
\begin{equation}
L_\beta = -i \frac{\partial}{\partial \alpha} +\beta
\end{equation}
Here, $\beta$ is a rational number equal to $p/q$; $p$ and $q$ are integers having no common divisors for a q-fold covering. The orthonormal basis vectors in the Hilbert space 
\begin{equation}
e_n(\alpha) = e^{i n \alpha} ~ n \in Z
\end{equation}
can be used to construct state vectors in the Hilbert space on the circle. The nontrivial topology arises due to the homotopy group of the nonsimply connected circle. The number $\beta$ defines the different irreducible representations in the same Hilbert space defined by the basis vectors (76). A unitary transformation on (76) can be made to define separate Hilbert spaces for each $\beta$. Now, the operator becomes independent of $\beta$
\begin{equation}
L =  -i \frac{\partial}{\partial \alpha}
\end{equation}
The basis vectors will have constant multiplicative phase factor $e^{i 2\pi \beta}$. The eigenvalues $(n+\beta)$ do not change under the unitary transformation. Modular angular momentum as a physical process for the Aharonov-Bohm phase based on this has been proposed in \cite{27}.

Since vortex model of photon is based on the fluid flow it is not same as the electron motion. To understand the role of E(2) group in the vortex model of photon a critical re-examination of electron motion in 2D plane is necessary. Free electron motion has unambiguous symmetry group of translations along x-axis and y-axis, and rotation around z-axis. In the case of uniform magnetic field along z-axis the cyclotron motion in the xy-plane has the central extension of the Euclidean group $\tilde{E}(2) $ as the symmetry group \cite{28}. Martin's model for the Aharonov-Bohm effect considers two cases: (i) electron motion has 1-parameter translation groups in x-direction and y-direction, and 1-parameter rotation group around z-axis when there is no current flow in the solenoid encircled by the electron. The symmetry group is E(2). And, (ii) if the current is switched on in the solenoid, and the solenoid is mathematically represented as a flux line then the electron motion encircling the flux line has rotational symmetry but translational symmetry is broken. In this case, the suggested symmetry group is the q-fold covering group. The freedom to change the solenoid current and the flux results into the phase shift dependent on the magnetic flux. Martin does not consider electron motion that does not encircle the flux line. In this case, there will be a phase shift in addition to the phase arising from the path difference, and it can be compensated by a gauge transformation. This Aharonov-Bohm phase shift, unlike Aharonov-Bohm effect, has no observable effect and the symmetry group is E(2). In the original paper \cite{29} the remarkable ingenuity seems to be that of distinguishing geometric effect, i. e. Aharonov-Bohm phase shift, and topological effect, i. e. Aharonov-Bohm effect. 

In the case of vortex line for photon there is no freedom to change the  vortex strength, i. e. the  analogue of magnetic flux in the Aharonov-Bohm effect. Note that the hole where curl of the vector field is non-zero in the fluid flow has intrinsic attribute. One way to understand the intrinsic nature of the topological obstruction is to calculate the angular momentum using (54)-(55)
\begin{equation}
L_z = x p_y - y p_x = x(\hbar \mathcal{A}_y) -y  (\hbar \mathcal{A}_x) =\hbar
\end{equation}
Since $-(\mathcal{A}_x, \mathcal{A}_y)$ is also a vortex solution, the angular momentum for this case is $-\hbar$. The transverse fields in the $\mathcal{A}$-vector waves give the intrinsic spin of photon. Obviously one cannot have Euclidean group representation based on either (75) or (77) for photon as spin has fixed value. The interpretation of the Euclidean group in the singular vortex fluid flow is sought in terms of the number of vortex filaments  assuming that it is defined by the de Rham period (58).  Let us interpret (58) as quantized vorticity, then a possible situation is that the vortex filaments are uniformly distributed on the 2D plane. Thus $n_R$  photons with spin $ \hbar$ and $n_L$ photons with spin $-\hbar$ constitute the polarized light beam with spin angular momentum equal to $(n_R -n_L) \hbar$. In this case, we have translational symmetry along x-axis and along y-axis for uniformly spaced vortex lines, i. e. photons. The whole system also has 1-parameter rotational symmetry about z-axis. The Euclidean group E(2) would arise as a natural symmetry group. 

\section{\bf Conclusion}

The main results of the present work are as follows. Pure vector field description for a single photon is established independent of the Maxwell fields. Considerations on the nature of various kinds of topological objects shows that the singular vortex line is the most suitable representation of a physical photon: spin is an invariant topological property, vorticity is defined by de Rham period, and the longitudinal wave propagation carrying the topological defect gives wave and particle attributes to photon. 

There seem to be two directions for developing the photon model. First, to derive the electromagnetic fields and the Maxwell equations based on $\mathcal{A}^\mu$ there are two possibilities. Instead of the uniformly distributed and equally spaced vortices discussed in the preceding section a scalar function interpreted as the number density of vortex lines (photons) $\rho_p$ the vector field $\rho_p \mathcal{A}^\mu$ can be constructed to define $F^{\mu\nu}$. One can also have a more general framework in which the vector field has two parts
\begin{equation}
A^\mu = A^\mu_{nonsingular} + \rho_p \mathcal{A}^\mu
\end{equation}
This framework would be similar to the two-component fluid theory of superfluids \cite{30}.

The second direction is to seek alternative to the quantum theory based on topological quantization and stochastic mechanics. Though Post made significant contribution on de Rham periods and quantization \cite{14} his repetitive and non-rigorous criticisms on the Copenhagen interpretation of quantum mechanics obscured his proposition of quantum cohomology \cite{31}. We believe topology and stochasticity have a potential to offer strong viable alternative to quantum mechanics.

{\bf APPENDIX}

An elementary brief introduction to the $\vee$-representation of the Clifford algebra in 4D space-time following \cite{10} is given in this appendix. Denoting differential 1-form as $\sigma^\mu =dx^\mu$, the Cartan exterior product gives 16 basis p-forms
\begin{equation}
1, \sigma^\mu , \sigma^\mu \wedge \sigma^\nu , \sigma^\mu \wedge \sigma^\nu \wedge \sigma^\lambda, \omega^4 = \sigma^1 \wedge \sigma^2 \wedge \sigma^3 \wedge \sigma^4
\end{equation}
Here, $\mu \neq \nu \neq \lambda$, and $\omega^4$ is 4-volume element.  Metric using  a scalar form is
\begin{equation}
g^{\mu\nu} =(\sigma^\mu , \sigma^\nu )
\end{equation}
An associative multiplication $\vee$ between the basis p-forms using the Cartan exterior product and the contractions is defined. For example,
\begin{equation}
\sigma^\mu \vee \sigma^\nu = g^{\mu\nu} + \sigma^\mu \wedge \sigma^\nu
\end{equation}
One can define antisymmetric tensor fields of rank $k$ as tensor type $f_k$: these are expanded onto a basis of differential forms (79) but given the $\vee$ product. The usual 4-vector $a_\mu$ denoted by $a$ has the scalar product $(a, b) = a_\mu b^\mu$.

Holomorphic condition for a tensor $f$ is
\begin{equation}
D \vee  f =0
\end{equation}
where the differential operator $D$ can be written with the $\sigma$-basis  as $\partial_\mu \sigma^\mu$, and $D f = D \vee F$, $F$ being the usual tensor. In 2D the condition (82) leads to the Cauchy-Riemann equations. In 4D space-time, $D= -{\bf \nabla}+ \partial^4 \sigma^4$. For a vector $a$ we have
\begin{equation}
D \vee a = {\bf \nabla}.{\bf a} +\partial^4 a^4 + (-\partial^4 {\bf a} - {\bf \nabla} a^4) \wedge \sigma^4 +\omega^3 \vee {\bf \nabla} \times {\bf a}
\end{equation}
The holomorphic condition $D \vee a =0$ in this case gives $\partial_\mu a^\mu =0; ~ \partial^\mu a^\nu -\partial^\nu a^\mu =0$. Thus we get the set of field equations (3), (4) and (5).


\begin{thebibliography}{99}
\bibitem{1} A. Pais, Some Strangeness in the Proportion, Edited by Harry Woolf (Reading MA: Addison-Wesley) p. 219
\bibitem{2} W. E. Lamb, Jr. Anti-photon, Appl. Phys. B66, 77 (1995)
\bibitem{3} L. Mandel and E. Wolf, Optical Coherence and Quantum Optics (C. U. P. 1995)
\bibitem{4} S. C. Tiwari, Photons and vortices, J. Mod. Opt. 46, 1721 (1999)
\bibitem{5} S. C. Tiwari, Relativity, entanglement and the physical reality of the photon, J. Opt. B: Quantum Semiclassical Opt. 4, S39 (2002)
\bibitem{6} S. C. Tiwari, Photon spin, zero-point energy and black-body radiation, arXiv: quant-ph/0606031 v1
\bibitem{7} S. C. Tiwari, Nature of the angular momentum of light: rotational energy and geometric phase, arXiv:quant-ph/0609015v1
\bibitem{8} S. C. Tiwari, Topological photon, J. Math. Phys. 49, 032303 (2008)
\bibitem{9} J. R. Oppenheimer, Note on light quanta and the electromagnetic field. Phys. Rev. 38, 725 (1931)
\bibitem{10} N. Salingaros, Electromagnetism and the holomorphic properties of spacetime, J. Math. Phys. 22, 1919 ((1981)
\bibitem{11} D. Han, Y. S. Kim, and D. Son, Photon spin as a rotation in gauge space, Phys. Rev. D 25, 461 (1982)
\bibitem{12} S. C. Tiwari, Superluminal Phenomena in Modern Perspective Faster than Light Signals: Myth or Reality? (Rinton Press, NJ, 2003)
\bibitem{13} R. H. Good, Jr. , Particle aspect of the electromagnetic field equations, Phys. Rev. 105, 1914 (1957) 
\bibitem{14} E. J. Post, Quantum Reprogramming (Kluwer, Dordrecht, 1995)
\bibitem{15} J. F. Nye, Polarization effects in the diffraction of electromagnetic waves: the role of disclinations, Proc. R. Soc. (London) A 387, 105 (1983)
\bibitem{16} M. Jammer, The Philosophy of Quantum Mechanics (Wiley, N. Y. 1974)
\bibitem{17} L. de Broglie, A new interpretation concerning the coexistence of waves and particles, in Perspectives in Quantum Theory, Edited by W. Yourgrau and A. van der Merwe (MIT Press, 1971)
\bibitem{18} E. Whittaker, A History of the Theories of Aether and Electricity (Thomas Nelson, 1951)
\bibitem{19} M. S. Soskin and M. V. Vasnetsov, Singular optics, Chapter 4, p 219 in Progress in Optics, vol. 42, 2001, Editor E. Wolf (Elsevier, North Holland)
\bibitem{20} R. Kiehn, Private communication; Reprinted in S. C. Tiwari, The Learning Mind: Digital Revolution or Thought Revolution?
E-book 11753, pothi.com (ResearchGate)
\bibitem{21} E. P. Wigner, On unitary representations of the inhomogeneous Lorentz group,  Ann. Math. 40, 149 (1939)
\bibitem{22} Y. S. Kim and E. P. Wigner, Cylindrical group and massless particles, J. Math. Phys. 28, 1175 (1987)
\bibitem{23} D. Han and Y. S. Kim, Little group for photons and gauge transformations, Am. J. Phys. 49, 148 (1981)
\bibitem{24} S. Weinberg, Photons and gravitons in S-matrix theory: Derivation of charge conservation and equality of gravitational and inertial mass, Phys. Rev. 135, B1049, (1964)
\bibitem{25} R. M. Kiehn, Periods on manifolds, quantization, and gauge, J. Math. Phys. 18, 614, (1977)
\bibitem{26} C. Martin, A mathematical model for the Aharonov-Bohm effect, Lett. Math. Phys. 1, 155 (1976)
\bibitem{27} S. C. Tiwari, Physical reality of electromagnetic potentials and the classical limit of the Aharonov-Bohm effect, Quantum Studies: Math. Found. 5, 279 (2018)
\bibitem{28} S. C. Tiwari, Symmetry of Dirac two-oscillator system. gauge invariance, and Landau problem, To appear in Mod. Phys. Lett. A; arXiv: 2306.01017 
\bibitem{29} Y. Aharonov and D. Bohm, Significance of electromagnetic potentials in the quantum theory, Phys. Rev. 115, 485 (1959)
\bibitem{30} R. P. Feynman, Application of quantum mechanics to liquid Helium, Prog. in Low Temperature Phys. vol. I, 17 (1955)
\bibitem{31} S. C. Tiwari, Book Review: Quantum Reprogramming by E. J. Post, Phys. Essays 7, 504 (1994) 
\end{thebibliography}
\end{document}